\documentclass[12pt]{iopart}

\usepackage[normalem]{ulem}
\usepackage{graphicx}
\usepackage[usenames]{color}
\def \dddot#1{{\stackrel{\ldots}{#1} }}
\def \xrightarrow#1{{\mathop{\,\longrightarrow\hspace{0.3cm}}_{#1}}}

\def \notag{\nonumber}
\def \text#1{{\mbox{#1}}}

\newcommand{\by}{{\bf y}}
\newcommand{\bx}{{\bf x}}
\newcommand{\bp}{{\bf p}}
\renewcommand{\br}{{\bf r}}

\newcommand{\be}{\begin{equation}}
\newcommand{\ee}{\end{equation}}

\newcommand{\beq}{\begin{eqnarray}}
\newcommand{\eeq}{\end{eqnarray}}

\def\H1{\widehat{H}_1}

\renewcommand{\e}{\ensuremath{\mathrm{e}}}

\def \figwa{0.42}

\def \etal{{\it {et al.}}}

\begin{document}
\title{Scaling approach to quantum non-equilibrium dynamics of many-body systems}

\author{Vladimir Gritsev$^{1}$}
\author{Peter Barmettler$^{1,2}$}
\author{Eugene Demler$^3$}
\address{$^1$Physics Department, University of Fribourg,
Chemin du Mus\'ee 3, 1700 Fribourg, Switzerland\\$^2$CPHT, \'Ecole Polytechnique, 91128 Palaiseau cedex, France
\\$^3$Lyman
Laboratory of Physics, Physics Department, Harvard University, 17
Oxford Street, Cambridge MA, 02138, USA }
\ead{vladimir.gritsev@unifr.ch}
\begin{abstract}

Understanding non-equilibrium quantum dynamics of many-body systems
is one of the most challenging problems in modern theoretical
physics. While numerous approximate and exact solutions exist for
systems in equilibrium,  examples of non-equilibrium dynamics of
many-body systems that allow reliable theoretical analysis, are
few and far between. In this paper we discuss a broad class of
time-dependent interacting systems subject to external linear and
parabolic potentials, for which the many-body Schr\"{o}dinger
equation can be solved using a scaling transformation.
We demonstrate that scaling solutions exist for both local and nonlocal
interactions and derive appropriate self-consistency equations.
We apply this approach to several specific experimentally relevant examples of
interacting bosons in one and two dimensions. As an intriguing
result we find that weakly and strongly interacting Bose-gases expanding from a
parabolic trap can exhibit very similar dynamics.
\end{abstract}

\date{\today}
\maketitle

\section{Introduction}

Understanding time evolution of complex quantum systems, often in
the presence of strong correlations between constituent particles,
is crucial for solving many fundamental problems in physics, from
expansion of the early universe, to heavy ion collisions, to pump
and probe experiments in solids. New questions of dynamical
evolution arise in recently realized artificial quantum many-body
systems, such as ultracold atoms in optical potentials or photons in
media with strong optical nonlinearities. These systems are
only weakly coupled to external heat baths and have a limited life-time,
thus many experiments require interpretation in terms of coherent
quantum dynamics rather than properties of equilibrium states.
On the positive side, these systems allow remarkable control of
parameters and open exciting opportunities for doing controlled
experiments exploring non-equilibrium many-body dynamics.

In the realm of many-body physics low-dimensional systems have a
special place. They have dramatically enhanced quantum and thermal
fluctuations and exhibit most surprising manifestations of strong
correlations. Rigorous theorems provide strong constraints on long-
range order and often such systems cannot be analyzed using
mean-field approaches even at zero temperature. Nevertheless,
equilibrium properties are well understood using methods specific to
low dimensions, such as Coulomb-gas representation of vortices in
two dimensions or effective low energy descriptions of
one-dimensional systems including Luttinger liquid and sine-Gordon
models (see e.g. ref. \cite{giamarchi}). However, such analysis
cannot be straightforwardly extended to non-equilibrium dynamics.
Most equilibrium theories focus on the low-energy part of the
spectrum while non-equilibrium dynamics can couple degrees of
freedom at very different energy scales
\cite{rigol,richardson1,richardson2,richardson3,calabrese,hubbard,luttinger1,luttinger2,dmrg,dmrg1,dmrg2,zener,barmettler,H-M}.
It would be highly valuable to have examples of many-body dynamics
of low-dimensional strongly correlated systems amenable to an
unbiased analytical treatment. These examples could be used not only
for analyzing experimental systems, but also for testing theoretical
calculations utilizing effective models or approximations and for
checking validity of new numerical approaches. In this paper we
propose such a class of non-equilibrium quantum problems with
time-dependent Hamiltonians which allow for a scaling ansatz of
many-body wave functions.

Scaling solutions in quantum dynamics were first discussed in the
context of a single harmonic oscillator with a time-dependent
frequency \cite{LR,PZe,MMT,Se,Schuch,Fischer}. This problem can be reduced
to a time-independent one by properly rescaling space and time.
Scaling transformation of variables is possible due to the existence
of a dynamical symmetry generated by dynamical invariants of the
system \cite{MMT,Se}. There are also extensions of this approach to
single particle problems with potentials of the Coulomb and inverse
square type \cite{BK,Sutherland,PZe}. In the context of many-body problems,
scaling has first been used within mean-field approaches to bosonic systems, for the classical Gross-Pitaevskii equation~\cite{Castin,KSS,Ghosh,son-wingate,tdGPE}. Beyond these effective one-body problems, scaling solutions exist for hard-core bosons in one dimension \cite{MG} and in the unitary limit of fermionic gases with infinite scattering length \cite{WC}; these are problems for which the interaction enters a constraint on the wave function of an otherwise non-interacting system analysis.
Away from these specific limits, Pitaevskii and Rosch~\cite{PR} introduced a
scaling ansatz for a two-dimensional many-body system of
particles interacting with contact or inverse square interaction and related the existence of such solution to a hidden $SO(2,1)$ symmetry. 
In this paper we further extend full many-body scaling solutions to more general types of interaction and arbitrary dimensionality. This generalization can be achieved by allowing three parameters of the system -- the mass, the interaction constant and the external potential -- to be time-dependent. Scaling solution is possible when the interdependence of these parameters is given by an Ermakov type equation, similar to the one discussed in earlier approaches \cite{MG,WC,PR}, and an additional self-consistency equation which depends on dimensionality of the system and the nature of interactions.

Dynamical control over the system parameters is possible in recently developed artificial quantum systems, such as trapped ultracold atomic gases, where the
effective interaction can be tuned using either Feshbach resonances
or by changing the transverse confining potential, whereas the
effective mass can be changed by application of the weak optical
lattice \cite{bloch-2008}. Also with photons in nonlinear optical devices, where the time-dependent
dispersion and Kerr nonlinearity can be achieved using
electromagnetically induced transparency
\cite{fermphot,fermphot1,fermphot2,fermphot3,meystre-1999}. In this
paper we propose applications of the scaling ansatz which are
experimentally relevant in the context of both of these systems.

We emphasize that apart from the tunability of the parameters no specific restrictions on the system properties are imposed.
Particles can obey fermionic, bosonic or mixed
statistics, interact by pairwise interaction, and be subject
to parabolic confining potential, to a linear potential, and to a
complex chemical potential. 
The basic idea of the scaling solution presented hereafter is to map the
non-equilibrium equations of motion to an equilibrium many-body
Schr\"odinger equation. The mapping is based on scaling functions
which relate correlation functions of time-dependent systems to
correlation functions of systems in equilibrium. Hence several
results known for equilibrium many-body systems can be directly
translated to non-equilibrium situations. The reverse conclusion is also true: from a measurement of the system out of equilibrium, e.g. a quantum gas after expansion, we can deduce its initial (equilibrium) properties \cite{LoG}.

The paper is organized as follows. In section \ref{sec:general} we
introduce a general formalism of scaling transformation for a
many-body Schr\"{o}dinger equation. In section \ref{sec:md} as an
example of application of our approach we compute momentum
distributions for one- and two-dimensional bosonic gases with
contact interactions released from a parabolic trap. Further details
are given in the Appendices, where we also discuss relation of our
work to classical integrability of time-dependent bosonic systems
with contact interactions.

\section{Scaling transformation -- general approach}
\label{sec:general}

Our starting point is the many-body Schr\"odinger equation for $N$
interacting particles in $D$ dimensions,
\beq
\label{nush1}
\frac{\partial\Psi({\bf x}_{1},\ldots,{\bf x}_{N};t)}{\partial
t}=H(t)\Psi({\bf
x}_{1},\ldots,{\bf x}_{N};t),\\
H(t)=-\frac{1}{2m(t)}\sum_{i=1}^{N}\Delta_{x_{i}}^{(D)} -
\mu(t)N+{\bf g}(t)\sum_{i=1}^{N}{\bf
x}_{i}\nonumber\\
+\frac{m(t)\omega^{2}(t)}{2}\sum_{i=1}^{N}{\bf
x}_{i}^{2}+\sum_{i\neq j}V({\bf x}_{i}-{\bf
x}_{j};t),\notag
\eeq
where $\Delta_{x_{i}}^{(D)}$ is a $D$-dimensional Laplacian acting
on the coordinate ${\bf
x}_{i}=(x_{i}^{(1)},x_{i}^{(2)},\ldots,x_{i}^{(D)})$ of the particle
$i$  ($\hbar=1$ here). The external parameters (chemical potential
$\mu(t)$, linear potential ${\bf g}(t)$ and trapping frequency
$\omega(t)$) and the many-body interaction potential $V(\bx;t)$
depend explicitly on time. The chemical potential
$\mu(t)=\Re[\mu(t)]+i\Im[\mu(t)]$ can accommodate effects of
dissipation via its imaginary part \footnote{It is known that the
time evolution under non-Hermitian Hamiltonian in a spirit of
stochastic wave function description is equivalent to the
description of the open system by the Lindblad master equation, see
e.g. Ref. \cite{meystre-1999} }. While the dependencies on the linear
and chemical potentials can be removed by the Gallilei transformations and phase shifts respectively, we note that solving the quantum problem with
time dependence of the remaining parameters represents a
non-trivial task. For instances, unlike in the non-interacting case, the time dependence of the mass can not be removed by the simple redefinition of time variable.
 
We address the following question: under which conditions
Eq.~(\ref{nush1}) (the $\Psi$-system) can be transformed into the
Schr\"{o}dinger equation for a time-independent ($\Phi$-) system:
\beq
\label{nush2}
&&i\frac{\partial\Phi(\by_{1},\ldots,\by_{N};\tau)}{\partial
\tau}=H_0\Phi(\by_{1},\ldots,\by_{N};\tau),\\
&\!H_0&\!\!\!=-\frac{1}{2 m_0}\sum_{i=1}^{N}\Delta_{y_{i}}^{(D)}+\frac{m_0\omega_{0}^{2}}{2}\sum_{i}{\bf
y}_{i}^{2}+\sum_{i\neq
j}V_0(\by_{i}-\by_{j}).\notag
\eeq
We emphasize that so far in (\ref{nush2}) $\omega_{0}$ and $m_0$ are
unspecified parameters; in particular the $\Phi$-system can have
vanishing confining potential even when the $\Psi$-system is
confined. We assume that the time dependence of the pairwise
interaction potential enters through a single time-dependent
coupling $V({\bf x};t)\equiv V({\bf x})v(t)$ and
$V_0(\bx)=V(\bx)v_{0}$. We further assume that the interactions have
a scaling property and are characterized by the exponent $\alpha$,
which we take to be the same for both $\Psi$- and $\Phi$-systems,
\beq\label{intpot-scaling}
V(\lambda{\bf x})=\lambda^{\alpha}V({\bf x}).
\eeq
Most generic interaction potentials (or pseudo-potentials) satisfy a
scaling law (\ref{intpot-scaling}): $s$-wave interactions
$V_{s}(\bx)\propto\delta(\bx)$ ($\alpha=-D$), any algebraic law,
$V(\bx)\propto |\bx|^\alpha$, including Coulomb ($\alpha=-1$),
inverse square law ($\alpha=-2$) or dipole-dipole interactions
($\alpha=-3$). Other examples are ultracold fermions interacting via
$p$-wave channel which gives rise to the $\delta'$ pseudo-potential
($\alpha=D-1$). Also logarithmic potentials can be treated; scaling
of the logarithmic law produces a time-dependent shift to $\mu(t)$.

To express the solution of the time-dependent Schr\"odinger
equation (\ref{nush1}) in terms of the solution
$\Phi(\by_1,\ldots,\by_N;\tau)$ of the static equation (\ref{nush2})
we introduce the scaling ansatz
\beq\label{trans}
\Psi({\bf x}_{1},\ldots,{\bf
x}_{N};t)&=&e^{i[F(t)\sum_{i=1}^{N}{\bf
x}^{2}_{i}+{\bf G}(t)\sum_{i=1}^{N}{\bf
x}_{i}+M(t)N]}\notag\\&\times&\frac{1}{R^{N}(t)}\Phi(\by_1,\ldots,\by_N;\tau)\,,
\eeq
with ${\bf y}_{i}=({\bf x}_{i}/L(t))+{\bf S}(t)$ and
$\tau\equiv\tau(t)$.
Direct calculation shows (see \ref{app:scaling}),
that this ansatz is valid if the scaling functions $R(t), L(t), F(t),\tau(t), {\bf G}(t), {\bf S}(t),
M(t)$ satisfy a set of coupled differential equations,
\beq\label{system}
&\dot{R}(t)=&\frac{1}{m(t)}D F(t)
R(t)-\Im[\mu(t)]R(t),\label{1}\\
&\dot{L}(t)=&\frac{2}{m(t)}F(t)
L(t)\label{2}\\
&\dot{F}(t)=&-\frac{2}{m(t)}F^{2}(t)-\frac{m(t)\omega^{2}(t)}{2}+\frac{m_{0}^{2}\omega_{0}^{2}}{2L^{4}(t)m(t)},\label{3}\\
&\dot{\tau}(t)=&\frac{m_{0}}{m(t)L^{2}(t)},\label{4}\\
&\dot{M}(t)=&-\frac{{\bf
G}^{2}(t)}{2m(t)}-\Re[\mu(t)]+\frac{m_{0}^{2}\omega_{0}^{2}{\bf
S}^{2}(t)}{2m(t)L^{2}(t)}, \\
&\dot{\bf S}(t)=&-\frac{{\bf G}(t)}{m(t)L(t)},\\
&\dot{\bf G}(t)=&-\frac{2F(t){\bf G}(t)}{m(t)}-{\bf
g}(t)+\frac{m_{0}^{2}\omega_{0}^{2}{\bf S}(t)}{m(t)L^{3}(t)},\\
&L^{-(\alpha+2)}(t)&=\frac{m(t)}{m_{0}}\frac{v(t)}{v_{0}}.\label{5}
\eeq
It is not obvious a priori that equations (\ref{1}-\ref{5}) can be
satisfied simultaneously for any reasonable time-dependencies of system
parameters $m(t)$, $v(t)$, $\omega(t)$.
Our next goal is
to show that there is a number of non-trivial cases for which
equations (\ref{1}-\ref{5}) are consistent with each other.
First of all we note that equations (\ref{1}) and (\ref{2}) imply
that $R(t)=[L(t)]^{D/2}\exp(-\int_{0}^{t}\Im[\mu(t)]dt)$. In the
absence of dissipation ($\Im[\mu(t)]=0)$ this condition is
equivalent to the conservation of the norm of the wave function
under the scaling transformation.  Eq. (\ref{2}) allows to express
$F(t)$ via $L(t)$, $F(t)=\frac{m(t)}{2}\dot{L}/L$, which can the be
substituted into the Eq.~(\ref{3}). This leads to the differential
equation for $L(t)$,
\beq
\label{eq:l}
\ddot{L}(t)+h(t)\dot{L}(t)+\omega^{2}(t)L(t)=\frac{m_0^2\omega_{0}^{2}}{m^{2}(t)L^{3}(t)}\,,
\eeq
where $h(t)=m_0\dot{m}(t)/m(t)$. The term with the first derivative
can be removed by the change of variables $L(t)=\exp[B(t)]y(t)$ with
$\dot{B}(t)=-h/2$. For $y(t)$ we obtain
\beq\label{ermakov-eq}
\ddot{y}(t)+\Omega^{2}(t)y(t)=\frac{\omega_{0}^{2}}{y^{3}(t)}\,,
\eeq
where
$\Omega^{2}(t)=\frac{1}{4}h^{2}-\frac{1}{2}\dot{h}+\omega^{2}(t)$.
Eq.~(\ref{ermakov-eq}) is the celebrated Ermakov equation \cite{LA}
first discovered in 1880 \cite{Ermakov}. This equation has been used
primarily for tracking invariants of the time-dependent harmonic
oscillator. In \ref{app:ermakov} we show how one can use the
non-linear superposition principle to reduce Eq.~(\ref{ermakov-eq})
to the linear equation. Once $L(t)$ is known, the remaining set of
equations for $S(t),M(t),{\bf G}(t)$ can be solved directly.

In summary, to find time-dependent parameters which admit a scaling
solution one can apply the following recipe: after specifying two
time-dependent functions $\omega(t)$ and $m(t)$ one obtains a
solution of the Ermakov equation (\ref{ermakov-eq}) from which one
determines  time-dependent interaction strength $v(t)$ consistent
with Eq. (\ref{5}). Solutions for the functions $M(t),{\bf
G}(t),{\bf S}(t)$ can then be obtained straightforwardly provided
that functions ${\bf g}(t)$ and $\mu(t)$ are explicitly specified.
Note that the complexity of our method (e.g. solving the Ermakov Eq.)
does not depend on the number of particles $N$.

The initial conditions for systems (\ref{nush1}) and (\ref{nush2})
are related to each other through Eq.~(\ref{trans}) applied at time
$t=0$:
\beq\label{init-cond}
\Psi({\bf x}_{1},\ldots,{\bf
x}_{N};0)&=&e^{i[F(0)\sum_{i=1}^{N}{\bf
x}^{2}_{i}+{\bf G}(0)\sum_{i=1}^{N}{\bf
x}_{i}+M(0)N]}\notag\\&\times&\frac{1}{R^{N}(0)}\Phi\left(\frac{{\bf x}_{1}}{L(0)},\ldots,\frac{{\bf x}_{N}}{L(0)};\tau(0)\right).\notag
\eeq
Generally  at $t=0$ the Hamiltonians controlling the dynamics of
$\Psi$- and $\Phi$-systems do not coincide. For example, they can
have different confining potentials, or one system can be in a trap
while the other one is in free space ($\omega_0=0$). In this paper
we focus on a finite initial trapping potential,
$\omega(0)=\omega_{0}>0$, for which we introduce the additional
assumption that at $t=0$ the two systems coincide. This means that
we have $m(t=0)=m_0$, $v(t=0)=v_0$ $F(t=0)=G(t=0)=M(t=0)=0$. At
$t>0$ the parameters of the $\Psi$-system begin to change in time
while the parameters of the $\Phi$-system remain constant. Since the
two systems coincide for $t<0$, the initial state of the
$\Psi$-systems at $t=0$ should correspond to the equilibrium state
of the $\Phi$-system. Existence of the scaling solution in one
dimension in the hard-core limit $v_0\rightarrow\infty$ has been
established previously \cite{MG}. Within our approach this can be
understood as follows: the first equation of  (\ref{5}) is trivially
satisfied, whereas other equations do not depend on the interaction
strength and remain valid. Another special case is the two-dimensional system with contact interactions studied previously by Pitaevskii and Rosch \cite{PR} ($D=-\alpha=2$), for which Eq. (\ref{5}) is satisfied by constant mass and interaction.

\section{Dynamics of Bose-gas with contact interaction released from the trap}
\label{sec:md}

In this section, {\it as an example}, we apply the scaling approach
to an ultracold Bose gas with contact interaction which is prepared
in a confined, weakly interacting initial state. The nontrivial
dynamics comes from a sudden switching off of the confining
potential from $\omega(t)=\omega_0$ at $t=0$ to $\omega(t)=0$ at
$t>0$. Solution of the scaling equation (\ref{eq:l}) for constant
mass $m(t)=m_0$, is then given by $L(t)=\sqrt{(1+\omega_0^2t^2)}$,
and consequently $F(t)=\frac{m_0\omega_0^2t}{2}/L^2(t)$.
In \ref{app:ermakovtrap} we examine additional scenarios
corresponding to varying mass which exhibit similar behavior of the
scaling functions. Here, we also assume that $\mu(t),{\bf g}(t)$ are
time-independent constants. 

To characterize the non-equilibrium dynamics it is convenient to
deal with correlation functions which can be easily derived within
the scaling approach (\ref{app:corr}). The dynamics of the
momentum distribution, for example, can be related to the
single-particle density matrix $g_{1}$ of the initial state,
\beq
n({\bf p},t)&=&[L(t)]^{D}\int_{-\infty}^{\infty} d{\bf x
}\int_{-\infty}^{\infty} d{\bf x'} g_{1}({\bf x},{\bf
x'};0)\notag\\\!\!&\times&\e^{-i [F(t)L^{2}(t)({\bf x}^{2}-{\bf x'}^{2})+L(t){\bf p}({\bf
x}-{\bf x'})]}\,.\label{momdistr}
\eeq
From the asymptotic behavior of the scaling functions
$L(t)\xrightarrow{\omega_0 t\gg 1} \omega_0 t$ and
$F(t)L(t)=m(t)\dot{L}(t)/2\xrightarrow{\omega_0 t\gg 1}
{m_0\omega_0/2}$ we can extract the long-time limit of the momentum
distribution using the stationary phase approximation (SPA),
\beq
n({\bf p},t)\xrightarrow{\omega_0 t\gg 1}\left(\frac{2\pi}{m(t)\dot{L}(t)}\right)^D
g_1(\frac{{\bf p}}{m(t)\dot{L}(t)},\frac{{\bf p}}{m(t)\dot{L}(t)};0)\,.\notag
\eeq
Hence the momentum distribution becomes fully determined by the {\it
density} distribution [$\rho(\bx,t)=g_1(\bx,\bx,t)$] of the initial
state.

For a quantitative description of dynamics we need to specify the
initial correlation function, which we take from earlier analysis of
effective theories for weakly interacting Bose-gases in harmonic
traps \cite{PGP,xia-2005,BMBT}. An important characteristic for a condensed state
with a sufficiently large number of particles is the  Thomas-Fermi
shape of the density profile,  $\rho({\bx})=\Theta(R_{TF}-|\bx|)(\mu/v_{0})\left(1-({\bx}/R_{TF})^2\right)$, where
$R_{TF}=\sqrt{2\mu/m_{0}}/\omega_0$ is the Thomas-Fermi radius.

\begin{figure}[ht]
 \centering
  \includegraphics[width=\figwa\textwidth,angle=0]{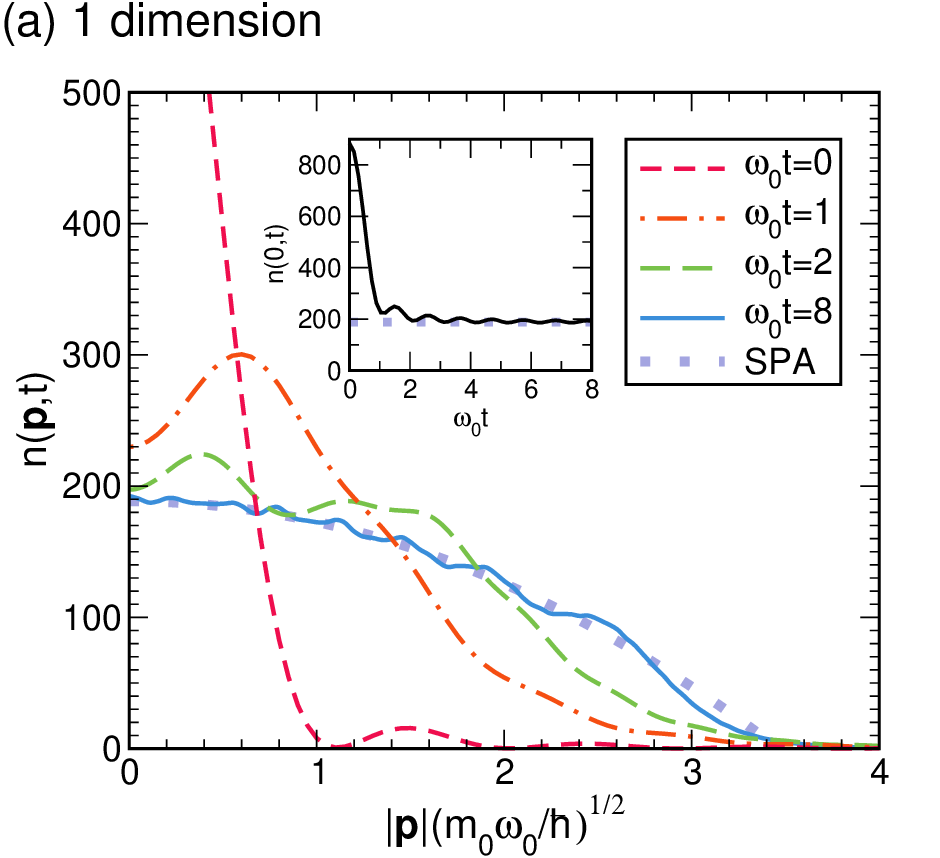}
  \includegraphics[width=\figwa\textwidth,angle=0]{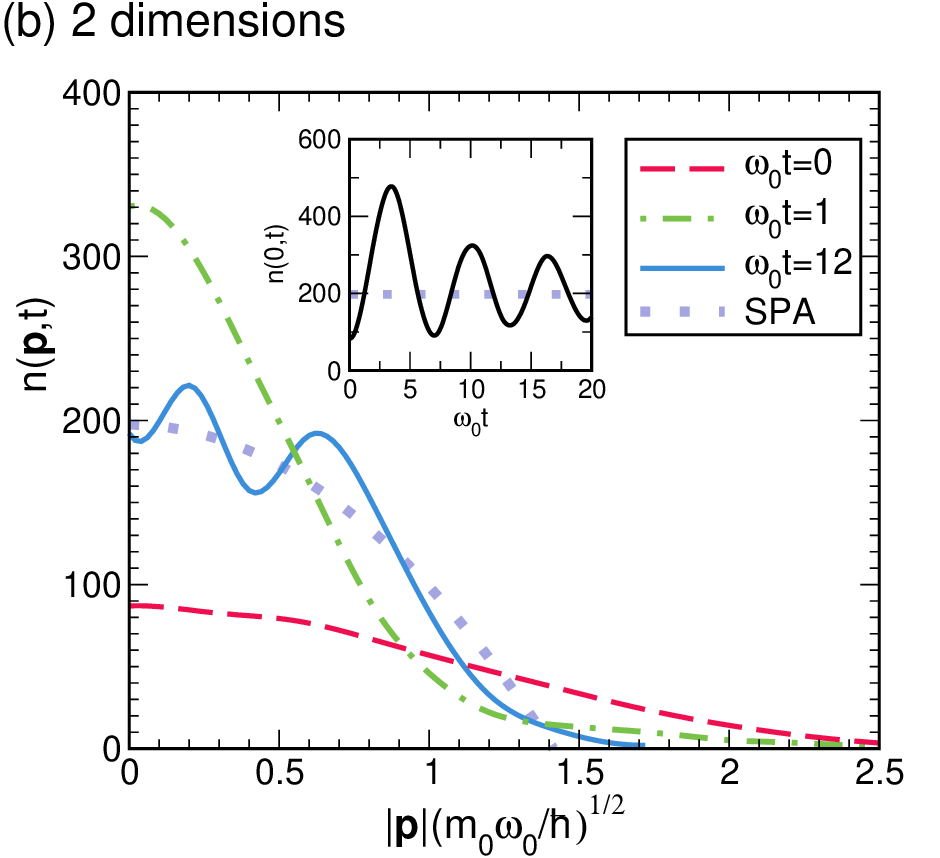}
  \caption{Temporal evolution of  momentum distribution functions
  following turning off the trap at $t=0$. The insets show the time
  evolution of the ${\bf p}=0$ component. The initial correlation
  functions are derived from effective theories (Refs. \cite{PGP,xia-2005,BMBT},
  see also  \ref{app:corr}). Dynamical evolution is obtained from numerical
  integration of Eq. (\ref{momdistr}). The stationary phase approximation
  (SPA) represents the asymptotic $t\rightarrow\infty$ result. Numerical
  errors are of the order of the line thickness. In the one-dimensional case
  (a) the system parameters are $N=140$, $k_BT=0.1\hbar\omega_0$,
  $v_{0}=0.2\sqrt{\hbar^3\omega_0/m_0}$, $R_{TF}=3.46\sqrt{\hbar/(m_0\omega_0)}$,
  $v(t)=v_0\sqrt{(1+\omega_0^2t^2)}$. In the two-dimensional case
  (b) the
  interaction strength is constant, $v(t)=v_0$ and  $N=16$, $k_BT=0.1\hbar\omega_0$,
  $v_{0}=0.2\hbar/m_0$, $R_{TF}=1.41\sqrt{\hbar/(m_0\omega_0)}$.\label{fig:npt}
  }
\end{figure}

First we analyze the one-dimensional case in the low-temperature
regime when the coherence length is of the order of the Thomas-Fermi
radius (Eq. (\ref{eq:g1trap1d}) of \ref{app:corr}).
According to the scaling equation (\ref{5}), for  contact
interactions, $V(\bx,t)=v(t)\delta(\bx)$ ($\alpha=-D$), the
interaction must be tuned inversely proportional to the scaling
function, $v(t)={v_0/L(t)}$. In Fig.~\ref{fig:npt}a results of
numerical evaluation of the momentum distributions  (\ref{momdistr})
for specific initial values are shown together with results from
SPA. The behavior of the ${\bf p}=0$ component is characterized by a
steep decay on a time scale $\omega_0^{-1}$ followed by slowly
dephasing oscillations, which are due to the finite extension of the
density profile and the quadratic phase factor in (\ref{momdistr}).
The corresponding period of oscillations $P$ is determined by the
Thomas-Fermi radius, $P\sim\frac{2\pi m_0}{\hbar R_{TF}^2}$.
Oscillations as a function of $|{\bf p}|$ at constant $t$ can  be
attributed to the finite Thomas-Fermi radius as well. Here the
quadratic phase factor leads to the oscillation period growing with
$|{\bf p}|$. In agreement with the SPA prediction, the momentum
distribution relaxes to a semi-circle law. This is remarkable, since
such a behavior has been previously associated with one-dimensional
Bose-systems in the \textit{strongly} interacting limit
($v_0\rightarrow\infty$) \cite{MG} only. In our case the interaction
strength is initially small and then even decreases in time. We note
that this can not be understood as effect of dilution due to
expansion of the system
because the effective one-dimensional interaction parameter
\cite{lieb-1963}, $\gamma\propto v(t)/\rho(t)\propto v(t)L(t)$,
remains constant.

In two dimensions $\alpha=-2$ and Eq. (\ref{5}) leads to
interactions which are constant in time. When the initial state is
weakly interacting (\ref{app:corr}), we choose an effective
theory which incorporates effects of quantum and thermal
fluctuations \cite{xia-2005}. Results of numerical evaluation of Eq.
(\ref{momdistr}) are shown in Fig.~(\ref{fig:npt}b). The momentum
distribution evolves very much like in the one-dimensional case and
is essentially determined by the initial density distribution and
the associated Thomas-Fermi radius. Here the number of particles
($N=16$) is set to be smaller than in the one-dimensional system.
Therefore the asymptotic stationary phase solution is  approached
slowly and oscillations dominate in the analyzed time window
$\omega_0 t\leq20$.
We checked that both in one and two dimensions the results are
robust against variation of temperature and interactions as long as
phase coherence is not destroyed.

The analysis of these examples leads to remarkable consequences. We
note that the stationary phase regime is reached  rather quickly
with momentum distribution determined by the initial density
distribution. Therefore specially designed initial density
distributions (equilibrium or not) can be used to {\it create}
specific momentum distributions, such as step-like fermionic ones,
on demand. It is remarkable that such behaviour,
which has been obtained previously in the strongly interacting limit,
persists down to arbitrarily weak strength of interaction. 
This is opposite to what is realized in time-of-flight experiments
of ultracold atoms released from a lattice \cite{GTF}, where the
expansion at sufficiently large times can be regarded as free and
momentum distributions get mapped to density profiles. By contrast
in our case we find that the real space density profile in the trap
determines momentum distribution after expansion (see Eq.
(\ref{momdistr})). While we do not discuss the appropriate time
evolution of $\omega(t)$, $m(t)$, and $v(t)$ here, we point out that
the time-of-flight  'far-field' limit \cite{GTF} may also be
captured formally by our scaling approach when the asymptotics of
$L(t)$ are linear and the contribution of the quadratic phase factor
in Eq. (\ref{momdistr}), $m(t)\dot{L}(t)$, vanishes in the long-time
limit. 
\section{Conclusions and outlook}

We used scaling ansatz to show that certain quantum non-equilibrium
problems with time-dependent parameters can be related to
equilibrium problems with constant parameters provided that the
time-dependent parameters satisfy a system of self-consistency
equations. This approach is valid for rather general types of
interactions and is not linked to the integrability of the model.
However, an integrable structure, when it exists, is consistent with
the scaling transformation. Solvability by the scaling ansatz is a
consequence of the non-relativistic dynamical symmetry which
received considerable attention recently in relation to the
non-relativistic version of AdS/CFT correspondence
\cite{ads1,ads2,ads3,ads4,ads5}. The appearance of this symmetry in
realistic many-body systems, which we discuss in this paper, can
open intriguing connections to the concept of AdS/CFT
correspondence.

We used scaling approach to analyze the problem of an abrupt
switching off of a confining potential for bosonic systems with
contact interactions in $d=1$ and $2$. Such experiments can be
performed using either ultracold atoms or photons in non-linear
medium. We find that the asymptotic momentum distribution is
essentially given by the initial density profile -- a phenomenon
which previously has been discussed only in the (Tonks-Girardeau)
limit of the infinitely strong repulsive one-dimensional Bose gas
\cite{MG}. Possible future applications of the scaling ansatz
include interaction quenches or transport phenomena (by considering
finite linear potentials). Extensions of our method to systems with
dissipation are also possible.

In our analysis we considered the situation when the scaling ansatz
is obeyed exactly. We expect however that our results remain
qualitatively valid even for systems with small deviations from the
exactly scalable Hamiltonians. For example, weak lattice potentials
should not have dramatic effects as long as the effective mass
approximation is applicable. Therefore one could achieve a full
description of time-of-flight experiments if the lattice potential
and interactions are tuned accordingly. Moreover it is conceivable
that on a phenomenological level the ansatz can be used even when
the time- and space-dependencies of system parameters do not fully
satisfy the consistency equations. The scaling solution could then
be seen as a universality class of non-equilibrium systems, very
much like a renormalization group fixed point at equilibrium. It
would be interesting to address this conjecture in experiments.

\section{Acknowledgements} We would like to thank D. Baeriswyl, I. Bloch, V. Cheianov, D.
Gangardt, M. Lukin, G. Morigi, A. Polkovnikov, M. Zvonarev for
useful discussions and remarks. This work is supported by DARPA,
MURI, NSF DMR-0705472, Harvard-MIT CUA and Swiss National Science
Foundation.

\appendix

\section{Derivation of the scaling equations}
\label{app:scaling}
We consider the ansatz (\ref{trans})
\beq
\Psi({\bf x}_{1},\ldots,{\bf
x}_{N};t)&=&\frac{1}{R(t)}\exp(i[F(t)\sum_{i=1}^{N}{\bf
x}^{2}_{i}+{\bf G}(t)\sum_{i=1}^{N}{\bf x}_{i}+M(t)])\notag\\
&\times&\Phi(\frac{{\bf
x}_{i}}{L(t)}+{\bf S}(t);\tau(t))
\eeq
for the transformation between the many-body Schr\"{o}dinger
equation with time-dependent parameters (Eq.~(\ref{nush1}) and the
equation (\ref{nush2}) with time-independent coefficients.
Calculating directly
\beq
\dot{\Psi}&=&(-\frac{\dot{R}}{R^{2}}+\frac{i\dot{F}}{R}\sum_{i=1}^{N}{\bf
x}_{i}^{2}+\frac{i\dot{{\bf G}}}{R}\sum_{i=1}^{N}{\bf
x}_{i}+i\frac{\dot{M}}{R})e^{i\phi(x_{i},t)}\Phi({\bf y}_{i},\tau))
\\&+&\frac{1}{R}e^{i\phi(x_{i},t)}\sum_{i=1}^{N}\frac{\partial\Phi({\bf y}_{i};\tau)}{\partial
{\bf y}_{i}}[{\bf x}_{i}(-\frac{\dot{L}}{L^{2}})+\dot{{\bf
S}}(t)]+\frac{1}{R}e^{i\phi(x_{i},t)}\frac{\partial\Phi({\bf
y}_{i};\tau)}{\partial \tau}\dot{\tau},\notag
\eeq
where for the sake of brevity we introduced $\phi({\bf
x}_{i},t)=F(t)\sum_{i=1}^{N}{\bf x}^{2}_{i}+{\bf
G}(t)\sum_{i=1}^{N}{\bf x}_{i}+M(t)$ and where the dot denotes the
derivative with respect to $t$, and
\beq
\frac{\partial\Psi({\bf x}_{i},t)}{\partial {\bf
x}_{i}}&=&\frac{1}{R}\left(2i F\sum_{i}{\bf x}_{i}
+{\bf G}\right)e^{i\phi(x_{i},t)}\Phi({\bf
	y}_{i},t)\\&+&\frac{1}{R}e^{i\phi(x_{i},t)}\frac{\partial\Phi({\bf
y}_{i},\tau)}{\partial{\bf y}_{i}},\notag\\
 \Delta_{x_{i}}^{(D)}\Psi({\bf
x}_{i},t)&=&\Big\{\left(\frac{2iFD}{R}+\frac{1}{R}(2iF{\bf
x}_{i}+i{\bf G})(2iF{\bf x}_{i}+i{\bf G})\right)\Phi({\bf y}_{i},\tau)\\
&+&\left(\frac{4iF{\bf x}_{i}+2i{\bf G}}{RL}\frac{\partial\Phi({\bf
y}_{i};t)}{\partial {\bf y}_{i}}+\Delta_{y_{i}}^{(D)}\Phi({\bf
y}_{i};t)\frac{1}{RL^{2}}\right)\Big\}e^{i\phi(x_{i},t)}.\notag
\eeq
Substituting this into the initial Schr\"{o}dinger equation
(\ref{nush1}) with time-dependent coefficients and adding and
subtracting the term $A(t)\sum_{i}{\bf x}_{i}^{2}$ with yet to be
determined function $A(t)$ we regroup the different contributions in
front of $\Phi({\bf y}_{i},\tau)$, $\partial\Phi({\bf
y}_{i},\tau)/\partial{\bf y}_{i}$, and $\Delta_{y_{i}}$. Each
group has several contributions proportional to ${\bf
x}^{0}_{i},{\bf x}_{i},{\bf x}_{i}^{2}$ which are linearly
independent and must be treated separately. This is how
conditions expressed by Eqs.(\ref{3}) appear. The remaining
equation has  the form of a Schr\"{o}dinger equation with
time-dependent coefficients
\beq
i\frac{\partial\Phi({\bf
y}_{i},\tau)}{\partial\tau}\dot{\tau}&=&-\frac{1}{2m(t)L^{2}(t)}\Delta_{y_{i}}\Phi(y_{i},\tau)\\
&+&\left[A(t)L^{2}(t)\sum_{i}{\bf
y}_{i}^{2}+L^{\alpha}(t)v(t)V({\bf y}_{i}-{\bf
y}_{j})\right]\Phi({\bf y}_{i},\tau).\notag
\eeq
We note that to compensate the terms appearing after the change
${\bf x}_{i}\rightarrow {\bf y}_{i}$ in the quadratic potential we
get terms proportional to $\omega_{0}^{2}$ in the
Eqs.~(\ref{2}-\ref{5}). Now, requiring that the three unknown
functions $\tau, L(t), A(t)$ satisfy
\beq\label{3-eq}
\dot{\tau}=\frac{m_{0}}{L^{2}(t)m(t)},\qquad
v_{0}\dot{\tau}=v(t)L^{\alpha}(t),\qquad
A(t)L^{2}(t)=\dot{\tau}\frac{m_{0}\omega_{0}^{2}}{2}
\eeq
we obtain the remaining conditions in the set of
Eqs.(\ref{1}-\ref{5}). Under this conditions the Schr\"{o}dinger
equation for the function $\Phi(y,\tau)$ has no time-dependent
coefficients. From the conditions (\ref{3-eq}) above we determine
the function
\beq
A(t)=\frac{m_{0}\omega_{0}^{2}\left[(v(t)m(t)\right]^{\frac{4}{\alpha+2}}}{2m(t)v_{0}^{\frac{4}{\alpha+2}}}
\eeq
Therefore we find that when pairwise potentials obey
Eq.~(\ref{intpot-scaling}), and the systems of Eqs
(\ref{1})-(\ref{5}) is satisfied, Eq.~ (\ref{nush1}) is indeed
mapped to Eq.~(\ref{nush2}).

\section{Analysis of the scaling equations and their solutions -- the Ermakov equation and dynamical symmetry}
\label{app:ermakov}
\subsection{General properties of the Ermakov
and related equations}

In this Appendix we briefly overview some general properties of the
Ermakov (sometimes spelled as Yermakov) equation which plays such a fundamental role in
our formalism. We also point out the relation of this equation with
the Riccati equation and with the linear differential equation with
variable coefficients. The Riccati equation directly appears in our
approach in some limiting cases.

The Ermakov \cite{Ermakov} equation is defined as follows
\beq
\ddot{y}(t)+f(t)y(t)=\frac{a}{y(t)^{3}}.
\eeq
Here $a$ is some $t$-independent constant. If there is a nontrivial
solution of the second order differential equation
\beq
\ddot{x}(t)+f(t)x(t)=0
\label{eq:ermakovhom}
\eeq
then the transformation
\beq
\xi(t)=\int_{0}^{t}\frac{d\tau}{x^{2}(\tau)},\qquad z=\frac{y}{x}
\eeq
puts the Ermakov equation into the form
\beq\label{reduced}
z_{\xi\xi}=az^{-3}.
\eeq
where the subscript denotes the derivative. The solution for the
initial equation then follows immediately
\beq
C_{1}y^{2}=ax^{2}+x^{2}(C_{2}+C_{1}\int\frac{dt}{x^{2}})^{2}
\eeq
where $C_{1,2}$ are arbitrary constants. If we take two solutions of
the linear (Hill) equation to satisfy initial data $x_{1}(0)=x_{1}$,
$\dot{x}_{1}(0)=\dot{x}_{1}$ while $x_{2}(0)=0$, $\dot{x}_{2}\neq 0$
then a general solution of the Ermakov equation is given by a
nonlinear superposition principle,
\beq
y(t)=\sqrt{x_{1}^{2}(t)+\frac{1}{w^{2}}x_{2}^{2}(t)}
\eeq
where $w=x_{1}\dot{x}_{2}-x_{2}\dot{x}_{1}$ is a constant Wronskian.

Now, provided the linear equation for $x(t)$ is satisfied, the
function $u(t)$ defined as
\beq
x(t)=\exp(-\int_{0}^{t}u(t)dt)
\eeq
satisfies the Riccati equation,
\beq
\dot{u}-u^{2}=f(t)
\eeq

This demonstrates that all three equations are closely related:
Ermakov, linear second order differential equation with variable
coefficients and the Riccati equation. Other remarkable equations
are also connected to the Ermakov equation. For example (taking
$a=1$ for simplicity in (\ref{reduced})) and defining
$\xi(t)=z(t)^{-2}$ we obtain
$\xi\ddot{\xi}-(3/2)(\dot{\xi})^{2}+2\xi^{4}=0$. Now, defining
$w(t)$ via $\xi(t)=\alpha \dot{w}/w$ with $\alpha^{2}=-1/4$ we
obtain a Kummer-Schwarz equation
$\dot{w}\dddot{w}-(3/2)(\ddot{w})^{2}=0$.

In some limiting situations (e.g. $\omega_{0}=0$, see the next
appendices) the Riccati equation appears naturally in our approach,
so we sketch some of its properties here. The general Riccati
equation with time-dependent coefficients
\beq\label{ric}
\dot{u}(t)=f(t)u^{2}(t)+g(t)u(t)+h(t)
\eeq
can be transformed into the second order differential equation
\beq\label{sec}
f(t)\ddot{y}(t)-[\dot{f}(t)+f(t)g(t)]\dot{y}(t)+f^{2}(t)h(t)y(t)=0
\eeq
by the following substitution $y(t)=\exp(-\int f(t)u(t)dt)$. In many
cases a particular solution of (\ref{sec}) is easier to find than
the one for the (\ref{ric}).

The Riccati equation has a remarkable property: if there is a known
particular solution $u_{0}(t)$ of (\ref{ric}), then the general
solution of (\ref{ric}) is given by
\beq\label{sol}
u(t)&=&u_{0}(t)+\Phi(t)\left[C-\int f(t)\Phi(t)dt\right]^{-1}\\
\Phi(t)&=&\exp\left[\int(2f(t)u_{0}(t)+g(t))dt\right]
\eeq
where $C$ is an arbitrary constant. The particular solution
$u_{0}(x)$ corresponds to $C=\infty$.

The property (\ref{sol}) allows the construction of many solutions
of (\ref{ric}) for given functions $f(t),g(t),h(t)$. If, for
example, $f(t)=1$, $g(t)$ is arbitrary and $h(t)=-(a^{2}+ag(t))$ a
particular solution is $u_{0}(t)=a$, and a general solution is then
\beq
u(t)=a+\Phi(t)[C-\int\Phi(t)]^{-1},\quad \Phi(t)=\exp(2at+\int
g(t)dx)
\eeq
for arbitrary $C$. For example for $f(x)=1$, $g(x)=0$, $h(x)=bx^{n}$
we obtain
\beq
u(t)=-\frac{\dot{w}(t)}{w(t)}, \quad
w(t)=\sqrt{t}[C_{1}J_{\frac{1}{2k}}(\frac{1}{k}\sqrt{b}t^{k})+C_{2}Y_{\frac{1}{2k}}(\frac{1}{k}\sqrt{b}t^{k})],\\
k=\frac{1}{2}(n+2), \qquad \mbox{for}\qquad \qquad n\neq 2\\
u(t)=\frac{\lambda}{t}-t^{2\lambda}(\frac{t}{2\lambda+1}t^{2\lambda}+C)^{-1},
\qquad\mbox{for}\qquad n=-2,
\eeq
where $\lambda$ is a root of $\lambda^{2}+\lambda+b=0$.

\subsection{Relation to dynamical symmetry}

The Ermakov equation has the symmetry algebra isomorphic to
$sl(2,R)$, which is isomorphic to the algebra $so(2, 1)$ of
rotations on the surface of one-sheet hyperboloid. The property
(\ref{sol}) of the Riccati equation is related to the covariance of
the Riccati equation with respect to the fractional-linear
transformations which are generated by the action of $sl(2,R)$
algebra: the general solution can be expressed as a combination of
particular solutions. The same algebra (more explicitly, one of its
form, $su(1,1)$) appears as a dynamical symmetry of the quantum
harmonic oscillator, where Ermakov equation appears as well. This
has been first found in \cite{LR}. There a single quantum harmonic
oscillator with time-dependent frequency has been solved using the
methods of (adiabatic) invariants. An adiabatic invariant in this
case is a function of a solution of the Ermakov equation. This
approach has led to appearance of the Ermakov-Pinney type equation
\cite{Ermakov} in quantum mechanics (see e.g. \cite{Schuch} for a
recent review). In \cite{PZe} the same equation appears as a certain
consistency condition on the time-dependent rescaling of coordinate
and time in the wave function of the oscillator. It became clear
that these two approaches, one based on dynamical invariants and the
other on the scaling of dynamical variables, are equivalent. Indeed
the rescaling procedure can be regarded as a transformation,
generated by a certain symmetry group, i.e. $sl(2,R)$. The
generators of this symmetry are operators corresponding to dynamical
invariants. Therefore the successiveness of applicability of scaling
transformation implies the presence of {\it dynamical} symmetry
generated by the dynamical invariants \cite{MMT,Se}. For this
symmetry to hold one has to have a special class of potential terms
in the single-particle Hamiltonian \cite{BK}. Physically interesting
potentials correspond to the contact interaction, harmonic, Coulomb
and inverse square laws. That is why the scaling approach has been
applied to a Calogero-Sutherland model \cite{Sutherland} and {\it
classical} Gross-Pitaevski type systems
\cite{Castin,KSS,PR,Ghosh,tdGPE}. The appearance of the $su(1,1)$
dynamical symmetry in our non-relativistic systems suggests a
possible connection to non-relativistic version of the AdS/CFT
correspondence
\cite{ads1},\cite{ads2},\cite{ads3},\cite{ads4},\cite{ads5}. In fact
the Virasoro algebra of any conformal field theory contains
$su(1,1)$ as subalgebra.

\subsection{Specific solutions for $\omega_{0}>0$}

\label{app:ermakovtrap}
\begin{figure}[t]
 \centering
  \includegraphics[width=\figwa\textwidth,angle=0]{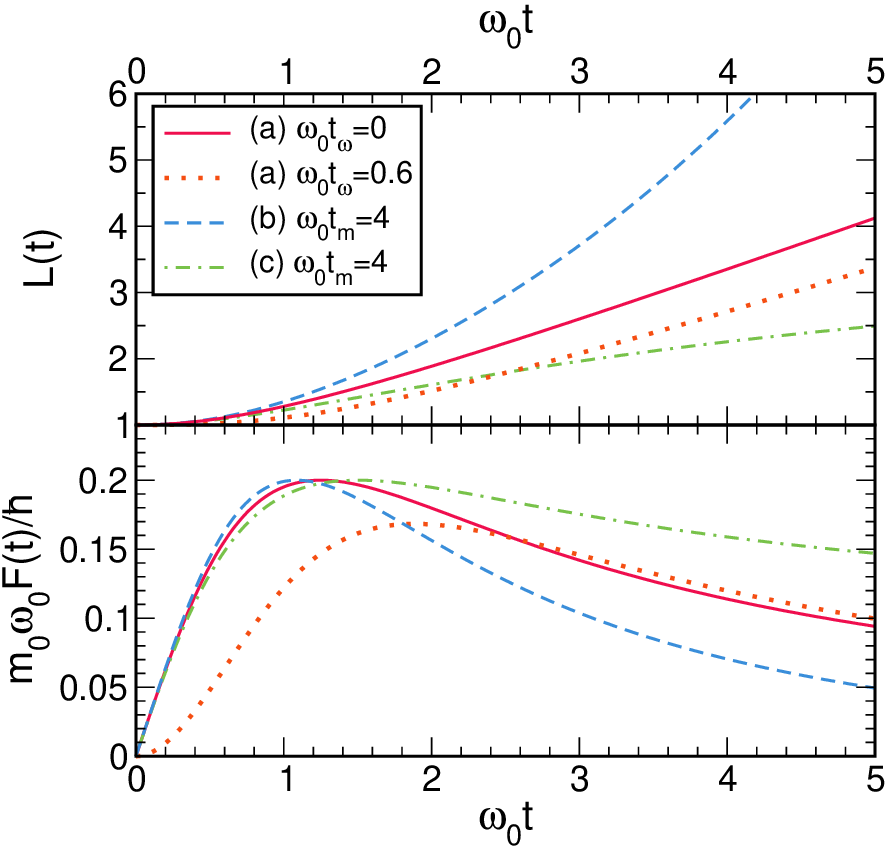}
  \caption{Scaling functions for $\omega_0>0$ ($\hbar$ reinserted by dimensional analysis).
  Each curve corresponds to one of the cases (a)-(c) analyzed in the text.\label{fig:scaling}}
\end{figure}

We compare examples for decreasing trapping potential and constant,
increasing and decreasing masses.
\begin{enumerate}
\item[(a)] \textit{Constant mass --} For the case of constant mass $m(t)=m_0$ we choose an
exponential decrease of the potential
$\omega(t)=\omega_0e^{-t/\tau_\omega}$. The two independent
solutions of the homogeneous equation (\ref{eq:ermakovhom}) read
$x_1(t)=J_0(2\tau_\omega\sqrt{\omega(t)})$,
$x_2(t)=Y_0(2\tau_\omega\omega_0\sqrt{\omega(t)})$. In fig.
\ref{fig:scaling} the resulting scaling functions obeying the
initial conditions $L(0)=1$, $F(0)=0$ are plotted. For sufficiently
small $\tau_\omega$ the functions are well described by the limit
$\tau_\omega\rightarrow 0$, for which the scaling solution reduces
to
\beq
L(t)=\sqrt{(1+\omega_0^2t^2)},\qquad
F(t)=\frac{m_0\omega_0^2t}{2}/L^2(t)\,.
\label{eq:scalinga}
\eeq

\item[(b)] \textit{Increasing mass --} We choose $m(t)=m_0e^{t/\tau_m}$ and, for sake of
simplicity, $\omega(t>0)=0$. The solution then reads
\beq
L(t)&=&\sqrt{1+(1-e^{-t/\tau_m})\tau_m^2m_0\omega_0^2},\notag \\
F(t)&=&(1-e^{-t/\tau_m}) \tau_m^2m_0\omega_0^2/L^2(t)\,,
\eeq
(plotted in fig.
\ref{fig:scaling}); this is similar to the scaling functions of
the case $(a)$, although the time is rescaled and in
the limit $t\rightarrow\infty$ the functions converge to the values
of the functions of case $(a)$ at $t=\tau_m$.

\item[(c)] \textit{Decreasing mass --} For $m(t)=m_0e^{-t/\tau_m}$ the scaling functions take the
form of case $(b)$ when replacing $\tau_m$ by $-\tau_m$ (see fig.
\ref{fig:scaling} for an illustration).
\end{enumerate}

We emphasize that the solutions do not depend on the dimensionality
of the system; only the  interaction constants, which have to
fulfill the consistency equation (\ref{5}), will do so.

i

\subsection{Specific solutions for $\omega_{0}=0$}
Based on two examples we demonstrate within our formalism, that if we
relate the non-equilibrium system in the trap to the system
without trap (the case $\omega_{0}=0$ in the main text) we directly
obtain a Riccati equation.

For $D=1$ Eq. \ref{5} reads $L(t)=m_0(m(t)c(t))^{-1}$
 (we define $c(t)=v(t)/v_0$) what we substitute in the equation for $L(t)$ to obtain
$F(t)=-(m(t)/2)\frac{d}{dt}\log[c(t)m(t)/m_0]$ . Consistency with the equation for $F(t)$ imposes
the following relation between three time-dependent parameters
\beq
-\frac{\dot{m}(t)}{2}\frac{d}{dt}\log[c(t)m(t)]&-&\frac{m(t)}{2}
\frac{d^{2}}{dt^{2}}\log[c(t)m(t)]\\&=&-\frac{m(t)}{2}(\frac{d}{dt}
\log[c(t)m(t)])^{2}-\frac{m(t)\omega(t)}{2}.
\eeq
By introducing $U(t)=\frac{d}{dt}\log(c(t)m(t))$
it reduces to the Riccati equation
\beq
\dot{U}(t)=\omega(t)-\frac{d}{dt}(\log[m(t)])U+U^{2}.
\eeq
The scaling ansatz (\ref{trans}) implies the relation between
initial conditions of the two systems:
$\Psi(t=0)=\exp(iF(0)\sum_{i}x_{i}^{2})\Phi(t=0)$ provided that
$L(t=0)=1$. The initial condition for the function $U(t)$ is not so
important for us because of the special property of the Riccati
equation, related to the B\"{a}cklund symmetry, which allows to
interrelate solutions with different initial conditions via a
rational function.

We note that the same equation describes the evolution of spin in a
time-dependent magnetic field. A general way to solve it is to
notice that under some change of variables it can be reduced to the
second-order liner differential equation
\beq
\ddot{u}-P(t)\dot{u}+Q(t)u=0, \,\,
P(t)=-\frac{d}{dt}\log[m(t)],\,\, Q(t)=\omega(t).
\eeq
Numerous explicit solutions are possible if we specify the
functions $\omega(t),m(t)$.

In the two-dimensional case we obtain from (\ref{1})-(\ref{5}) that
$R(t)\equiv L(t)$ and time-dependent parameters are connected by the
constraint $c(t)m(t)=c_{0}$. Then $F(t)=(m(t)/2)\frac{d}{dt}\log[L(t)]$.
Introducing $V(t)=\frac{d}{dt}\log L(t))$ and $h(t)=\frac{d}{dt}\log(m(t)/2)$ we
obtain
\beq
-\frac{dV(t)}{dt}=\omega(t)+h(t)V(t)+V^{2}(t)
\eeq
which is a Riccati equation for the coordinate scaling function
$L(t)$; its solution for given time-dependent parameters $m(t),
\omega(t)$ then defines a solution for the time-rescaling function
\beq
\frac{d\tau(t)}{dt}=\frac{m_0}{m(t)L^{2}(t)}
\eeq

To be specific we list two examples of dynamical parameters:

\begin{itemize}
 \item[(a)] \textit{Increasing mass --}  From the form of the Riccati equation it is
somewhat appealing to take $m(t)=m_0e^{\alpha t}$, and constant
$\omega(t)\equiv\Omega$. Then
\beq
c(t)=\phi(t)\exp[-\alpha t/2]
\eeq
where $\phi(t)=\sin(At+B)/C$ with $A,B,C$ related to
$\alpha$ and $\Omega$. In particular, for $m(t)=e^{2t}$, where $\Omega=1$,$A=B=C$ and $C\rightarrow0$ we
obtain $c(t)=(1+t)e^{-t}$.

\item[(b)] \textit{Constant mass --}  For $m(t)\equiv m_{0}$ the equation can
be transformed into the equation for the harmonic oscillator with
time-dependent-frequency $\omega(t)$ for which many known solutions
exist. Using these solutions we can extract the function $c(t)$. In
particular, for constant $\omega(t)=\Omega$ the solution for some
domain of parameters is
\beq
c(t)=\frac{1}{m_{0}\cos(\Omega t)}\,.
\eeq
In the simplest case of $m(t)=1$, $\omega(t)=0$ we obtain
$c(t)=-1/(1+t)$. This example is a many-body analogue of the
solution of the Hamiltonian with potential $V(x)=c(t)\delta(x)$
found in ref. \cite{BK} for a single-particle Schr\"{o}dinger
equation. Direct application of this solution can be found in the
ultracold Bose gas close to the confinement-induced resonance
\cite{Nagerl}.

\end{itemize}

Other examples of solutions of (\ref{ric}) can be found in the
literature, see e.g. ref. ~\cite{PZ}.

\section{Classical integrability of the nonlinear Schr\"{o}dinger
equation with time-dependent parameters}
\label{app:nse}
It is instructive to check whether the exact scaling transformation
we have studied in this paper is consistent with the property of
integrability of the nonlinear Schr\"{o}dinger equation (NSE). Here
we address this question for the classical NSE.

In the zero curvature representation, the NSE
\beq\label{nush}
i\frac{\partial\Psi}{\partial t}=-\frac{\partial^{2}\Psi}{\partial
x^{2}}+2 c |\Psi|^{2}\Psi
\eeq
is represented by the system of the first order differential
equations
\beq
\frac{\partial F}{\partial x}=U(x,t,\lambda)F,\qquad \frac{\partial
F}{\partial t}=V(x,t,\lambda)F,\qquad F=\left(\begin{array}{c}
                                          f_{1} \\
                                          f_{2}
                                        \end{array}\right)
\eeq
such that the matrices $U(x,t,\lambda)$ and $V(x,t,\lambda)$ which
depend on the spectral parameter $\lambda$ satisfy the condition
\beq\label{0curv}
\frac{\partial U}{\partial t}-\frac{\partial V}{\partial x} +[U,V]=0
\eeq
which is equivalent to the compatibility condition of the system,
\beq
\frac{\partial^{2}F}{\partial x\partial
t}=\frac{\partial^{2}F}{\partial t\partial x}
\eeq
and which is equivalent to the initial Schr\"{o}dinger equation. In
case of (\ref{nush}) one can establish that
\beq\label{uv}
U&=&U_{0}+\lambda U_{1},\qquad V=V_{0}+\lambda
V_{1}+\lambda^{2}V_{2}\\
U_{0}&=&\sqrt{c}(\bar{\Psi}\sigma_{+}+\Psi\sigma_{-}), \qquad
U_{1}=\frac{1}{2}i\sigma_{3}\\
V_{0}&=&ic|\Psi|^{2}\sigma_{3}-i\sqrt{c}(\frac{\partial\bar{\Psi}}{\partial
x}\sigma_{+}-\frac{\partial\Psi}{\partial x}\sigma_{-}),\qquad
V_{1}=-U_{0},\quad V_{2}=-U_{1}
\eeq
Conserved quantities are constructed from the matrices $U,V$ in a
known way. This method provides a direct way to various
generalizations of NSE. In particular one can obtain some
generalization where  the interaction parameter $c$ and the mass are
explicitly time-dependent functions. Introducing generalization of
(\ref{uv}) as
\beq
\tilde{U}&=&\left(
           \begin{array}{cc}
             -\frac{i}{2}\alpha(x,t) & \gamma(x,t)\bar{\Psi} \\
             \gamma(x,t)\Psi & \frac{i}{2}\beta(x,t) \\
           \end{array}
         \right),\notag\\
         \tilde{V}&=&\left(
                     \begin{array}{cc}
                       iA(|\Psi|^{2},\lambda(x,t)) & B(\bar{\Psi},\frac{\partial\bar{\Psi}}{\partial x},\mu(x,t)) \\
                       B^{*}(\Psi,\frac{\partial\Psi}{\partial x},\mu(x,t)) & -i D(|\Psi|^{2},\lambda(x,t)\\
                     \end{array}
                   \right)
\eeq
one can look for generalizations of integrable  NSE by appropriately
choosing the functions $\alpha(x,t),\beta(x,t),
\gamma(x,t),\lambda(x,t), \mu(x,t), A, B, D$. Analysis of the
zero-curvature condition (\ref{0curv}) in the case of inhomogeneous
time-dependent functions leads to a set of equations between those
functions and reveals a large class of solutions of the classical
equations of motions for NSE with time-dependent coefficients. To
get a consistency condition for a zero-curvature representation we
conclude that the spectral parameter should be an inhomogeneous
time-dependent function.

Some restricted form of this inhomogeneous time-dependent
$\tilde{U}-\tilde{V}$ pair has been considered in ref. \cite{Kundu}
where it was shown that a combination of space-time transformation
together with a $U(1)$ gauge transformation of the linear equations
for the $\tilde{U}-\tilde{V}$ pair and corresponding redefinition of
the field variables brings the system into the form of a homogeneous
time-independent NSE system, thus showing the integrability of a
time-dependent system. We note that a similar analysis has been
given in Ref. \cite{KDP}.

Although it is more difficult to show integrability on the quantum
level directly, presumably the property of integrability is not
violated in that case for specific choice of time-dependent
parameters which correspond to our scaling equations. A related
approach based on the inhomogeneity of spectral parameters for the
quantum sine-Gordon model has been recently presented in ref. \cite{Ku}.

\section{Scaling of correlation functions}
\label{app:corr}

With the scaling ansatz (\ref{trans}) the relation between the
single-particle correlation functions in the time-dependent and
time-independent systems is derived straightforwardly,
\beq\label{g1}
g_{1}^{(\Psi)}({\bf x},\bx',t)&=&
\int_{-\infty}^{\infty}\dots\int_{-\infty}^{\infty} d{\bf
x}_{2}\ldots
d{\bf x}_{N}\Psi^{*}({\bf x},{\bf x}_{2},\ldots,{\bf x}_{N};t)\notag\\
&&\qquad\times\Psi(\bx',{\bf x}_{2},\ldots,{\bf x}_{N};t)\notag\\
&=&\frac{1}{[L(t)]^{D}}g_{1}^{(\Phi)} \left(\frac{{\bf
x}}{L(t)},\frac{\bx'}{L(t)};0\right)\exp \left(-iF(t)({\bf
x}^{2}-\bx'^{2})\right).
\eeq
The labels in the $g_{1}$-function refer to the time-dependent
($\Psi$) and time-independent ($\Phi$) systems. From this expression
we can readily extract the density: $\rho^{(\Psi)}({\bf
x},t)=g_{1}({\bf x},{\bf x},t)=(1/L(t))\rho^{(\Phi)}({\bf
x}/L(t);0)$. The momentum distribution of a time-dependent system,
defined as
\beq
n^{(\Psi)}({\bf p},t)=\int_{-\infty}^{\infty} d{\bf x}
\int_{-\infty}^{\infty}d\bx' e^{-i{\bf p}({\bf x}-{\bf
y})}g_{1}^{(\Psi)}({\bf x},\bx',t),
\eeq
is then given by
\beq\label{momdistr2}
n^{(\Psi)}({\bf p},t)&=&[L(t)]^{D}\int_{-\infty}^{\infty} d{\bf x
}\int_{-\infty}^{\infty} d\bx' g_{1}^{(\Phi)}({\bf x},{\bf
y};0)\notag\\&&\exp[-i F(t)L^{2}(t)({\bf x}^{2}-\bx'^{2})-iL(t){\bf
p}\cdot({\bf x}-\bx')].
\eeq
Note that because of the quadratic term in the exponent the
integrations are nontrivial.

For the two-particle density matrix we find analogously
\beq
g_{2}^{(\Psi)}(x_{1},x_{2},x_{1}',x_{2}';t)&=&N(N-1)\int
dx_{3}\ldots
dx_{N}\Psi^{*}(x_{1},x_{2},\ldots,x_{N};t)\notag\\
&&\qquad\times\Psi(x_{1}',x_{2}',\ldots,x_{N};t)\nonumber\\
&=&\frac{1}{L(t)^{2}}g_{2}^{(\Phi)}\left(\frac{x_{1}}{L(t)},\frac{x_{2}}{L(t)},\frac{x_{1}'}{L(t)},\frac{x_{2}'}{L(t)};0\right)\notag\\
&&\qquad\times
\exp\left(-iF(t)(x_{1}^{2}+x_{2}^{2}-x_{1}^{'2}-x_{2}^{'2})\right).
\eeq
and  the  two-particle correlation function reads
\beq
\rho_{2}^{(\Psi)}(x,y;t)=g_{2}^{(\Psi)}(x,y,x,y;t)=\frac{1}{L^{2}(t)}\rho_{2}^{(\Phi)}\left(\frac{x}{L(t)},\frac{y}{L(t)};0\right).
\eeq

Other useful quantities such as non-equilibrium time-dependent
correlation functions (e.g. $n^{(\Psi)}(p,t,t')$ ) or the multi-mode
squeezing spectrum ($S(k,k';t,t')=\langle n^{(\Psi)}(p,t)
n^{(\Psi)}(p',t')\rangle$) can also be easily computed using the
scaling approach.


\section{Some technical details related to the derivation of 1D and 2D momentum distribution at equilibrium}
\subsection{Trapped weakly interacting Bose gases}
In order to describe a condensed Bose gas in a harmonic potential we
adopt results of previous works \cite{PGP,BMBT,xia-2005} which
consider phase fluctuations on top of the mean-field solution while
density fluctuations are assumed to be negligible. This is a valid
approximation for a sufficiently high number of weakly interacting
particles at low temperatures.  The temperature range where the
density fluctuations are suppressed is $T_{d}\gg T\gg T_{\phi}$
where the temperature of quantum degeneracy is
$T_{d}=N\hbar\omega_{0}$ and $T_{\phi}=T_{d}\hbar\omega_0/\mu$.

Generically, the single-particle correlation can be represented as
\beq
g_{1}({\bf x},\bx')=\sqrt{\rho({\bf
x})\rho(\bx')}\exp\left(-\frac{1}{2}\langle\left(\phi(\bx)-\phi({\bf
x}')\right)^{2}\rangle\right),
\label{eq:g1trap1d}
\eeq
where $\langle\phi(\bx)\rangle$ denotes the average over phase
fluctuations. We assume the validity of the Thomas-Fermi
approximation for the density
\beq
\rho({\bf x})\rightarrow \rho_{TF}({\bf
x})=\frac{\mu}{g}\left(1-(\frac{{\bf
x}}{R_{TF}})^{2}\right)\theta(1-|\frac{\bf x}{R_{TF}}|)\,,
\eeq
where $R_{TF}=\sqrt{2\mu/m_{0}}/\omega_0$ is the Thomas-Fermi
radius.

In a 1D geometry, taking into account thermal fluctuations and
neglecting contributions from quantum fluctuations, one obtains the
phase average \cite{PGP}
\beq
\langle\left(\phi(\bx')-\phi({\bf x})\right)^{2}\rangle =
\frac{4T\mu}{\hbar^{2}\omega^{2}}\left|\ln\left[\frac{(1-\frac{\bx'}{R_{TF}})(1+\frac{\bx}{R_{TF}})}{(1+\frac{\bx'}{R_{TF}})(1-\frac{\bx}{R_{TF}})}\right]\right|\,.
\eeq

For the 2D case an expression similar to the 1D case can be derived.
In this work we used the complete expression obtained by Xia et
\textit{al.} (Eq. (77) in Ref. \cite{xia-2005}), which explicitly
accounts for thermal \textit{and} quantum fluctuations. As a result,
at inter-particle distances much smaller than $2R_{TF}$ the
correlations decay exponentially with a decay rate approximately
given by $mk_{B}T/\left(2\pi\hbar^{2}\rho(0)\right)$. However, for
the dynamics studied in this paper we did not find significant
effects from quantum corrections.

\subsection{One- and two-dimensional uniform Bose gases}

For a one-dimensional Bose gas it was recently shown \cite{CC}  that
the effective field theory (Luttinger liquid) provides an extremely
accurate description for a single-body correlation function at
distances beyond the inter-particle separation. If we are not
interested in its large momentum behavior it is legitimate to use
this effective theory. The single particle correlation function in
time-independent theory is then well known (see e.g.
\cite{giamarchi}). For nonzero temperatures it is given by (we omit
oscillating terms)
\beq
g_{1}^{(\Phi)}(x,x';0)=\langle\Phi^{\dag}(x)\Phi(x')\rangle =
\rho_{0}B\left[\frac{\pi/\xi_{T}}{\rho_{0}\sinh(\pi(x-x')/\xi_{T})}\right]^{\frac{1}{2K}}
\eeq
where $\xi_{T}=\hbar v_{s}/T=\hbar^{2}\pi\rho/(m_0KT)$, $\rho_{0}$
is the uniform equilibrium density, $v_{s}$ is the sound velocity,
$K$ is a Luttinger parameter which is related to the interaction
strength $c$ and $B=(K/\pi)^{1/2K}$ is Popov's factor.

In the two-dimensional case, we consider a system below the
Berezinskii-Kosterlitz-Thouless (BKT) transition. The correlation
functions then decay algebraically with a temperature-dependent
exponent, which tends to the universal value $1/4$ when approaching
the BKT transition from below.


\section{Dynamics of initially uniform systems}
\subsection{Relating systems in the trap and without it}
\label{app:hom}
The scaling approach can be used to establish a relationship between
correlation functions in the model with time-dependent parameters
(the system $\Psi$) and the model with time-independent parameters
(the system $\Phi$). As we discussed in the main text, the trapping
frequency $\omega_{0}$ of the time-independent system is not fixed a
priory. In particular, it can be put equal to zero from the very
beginning. The scaling transformation therefore will relate the
system in the time-dependent trap and a uniform system. The set of
differential equations has to be modified accordingly. The aim of
this appendix is to look into the behavior of the momentum
distribution in this case.

The initial conditions state that the two wave functions are equal
at $t=0$. It means that the density distribution of the trapped
system is homogeneous, corresponding to the uniform one. This is
possible if we assume the existence of a length scale $l$ on which
this condition can be satisfied. Moreover we assume here that the
Thomas-Fermi radius of a trapped system is large enough such that
there is a finite region of  ${\bf x}\in [-l,l]^{D}$ where the
density is considered to be a constant. In  the absence of a
trapping potential this region is equal to the whole observation
area. We assume that this region is large enough to contain a
relatively large number of particles $N$. Using this length scale
$l$ as a sort of cut-off, we evaluate the momentum distribution in
the finite window $[-l,l]$ for examples of 1D and 2D systems at
finite temperature.

\subsection{Evaluation of momentum distributions in 1D for the uniform system}

In the Luttinger liquid approximation at finite temperature we
introduce $\xi_{\pm}=\pi(x-y)/\xi_{T}$ in terms of which
$g_{1}^{(\Phi)}(x,y;0)\sim(\sinh\xi_{-})^{-1/2K}$. This function
decays exponentially at large distances and the limits of
integration in $\xi_{-}$ domain can be therefore extended from
$[-l,l]$ to $(-\infty,\infty)$ to make analytic progress. One can
easily realize that because of the additional structure in the
exponent of Eq.~(\ref{g1}), the expression for the momentum
distribution is essentially different from the one at equilibrium.
The corresponding integral is
\beq
\int_{-\infty}^{\infty}\frac{e^{-iC(t)\xi_{-}}}{[\sinh|\xi_{-}|]^{1/2K}}&=&2^{-1+\frac{1}{2K}}
\Gamma(1-\frac{1}{2K})\\&&\quad\times\left(\frac{\Gamma[\frac{1}{4K}-i\frac{C(t)}{2}]}{\Gamma[1-\frac{1}{4K}-i\frac{C(t)}{2}]}+
\frac{\Gamma[\frac{1}{4K}+i\frac{C(t)}{2}]}{\Gamma[1-\frac{1}{4K}+i\frac{C(t)}{2}]}\right)
\label{Eq-1D}\notag
\eeq
where
$C(t)=F(t)L^{2}(t)\xi_{T}^{2}\xi_{+}/\pi^{2}+L(t)p\xi_{T}/\pi$. The
integration over $\xi_{+}$ is then performed in the finite interval
$[-l,l]$ corresponding to the size of the selected subsystem. The
expression (\ref{Eq-1D}) is proportional to the equilibrium momentum
distribution at $t=0$ provided that we take $L(0)=1$. We find
\beq\label{luttnp}
n(p,t)&=&2\rho_{0}L(t)\left(\frac{2K}{\xi_{T}\rho_{0}}\right)^{\frac{1}{2K}}
\Gamma(1-\frac{1}{2K})\notag\\
&\times&\int_{-l/L(t)}^{l/L(t)}d\xi_{+}\Big\{\frac{\Gamma[\frac{1}{4K}-\frac{i}{2}
(F(t)\bar{L}^{2}(t)\xi_{+}+p\bar{L}(t))]}\notag\\\notag&&\quad\quad\times
{\Gamma[1-\frac{1}{4K}-\frac{i}{2}(F(t)\bar{L}^{2}(t)\xi_{+}+p\bar{L}(t))]}\\&+&
\frac{\Gamma[\frac{1}{4K}+\frac{i}{2}(F(t)\bar{L}^{2}(t)\xi_{+}+p\bar{L}(t))]}
{\Gamma[1-\frac{1}{4K}+\frac{i}{2}(F(t)\bar{L}^{2}(t)\xi_{+}+p\bar{L}(t))]}\Big\}
\eeq
where $\bar{L}(t)=L(t)\xi_{T}/\pi$.

\begin{figure}[ht]
	\begin{center}
	\includegraphics[width=\figwa\textwidth,angle=0]{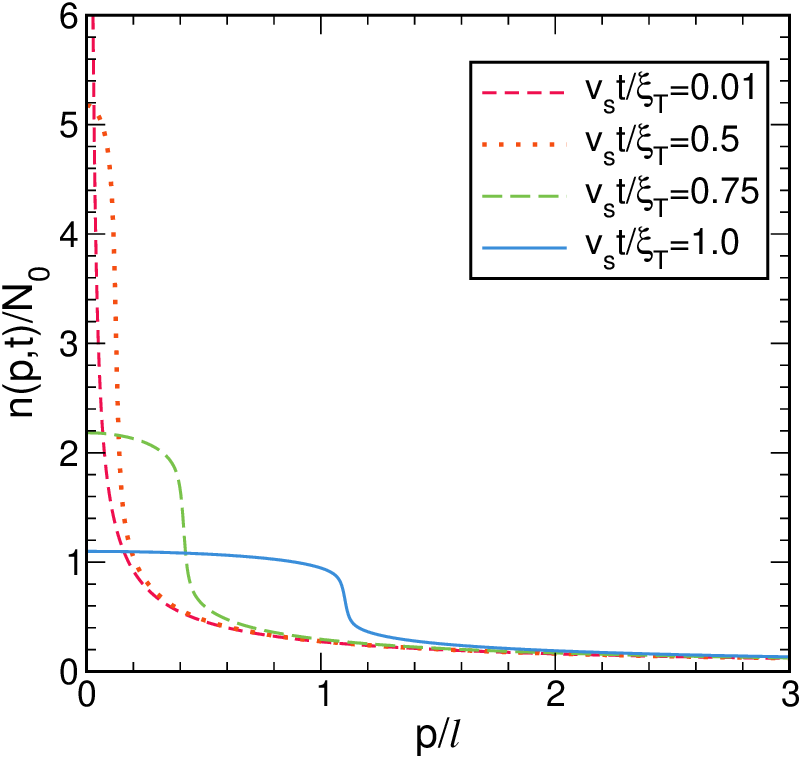}
\caption{The one-dimensional momentum distribution for different
times  computed within the Luttinger liquid model with $K=2$,
$l=10\xi_L$, normalized by a $K$- and $\rho_0$-dependent prefactor
$N_0=2\rho_{0}\left(\frac{2K}{\xi_{T}\rho_{0}}\right)^{\frac{1}{2K}}
\Gamma(1-\frac{1}{2K})$. The trap frequency is changing linearly,
$\omega(t)=t\left(\frac{v_s}{\xi_T}\right)^2$, while the mass varies
exponentially, $m(t)=m_0e^{2tv_s/\xi_T}$. }
\label{fermi}
\end{center}
\end{figure}

On the basis of this expression we have calculated a momentum
distribution for various particular functions $\omega(t)$ and
$m(t)$. Solving the set of consistency equations of Section
\ref{sec:general}, we obtained all the other functions $v(t), L(t),
F(t)$. This is illustrated in fig. \ref{fermi} for particular
choices of time-dependent functions $\omega(t), m(t)$ and
corresponds to a particular function $v(t)$ found from solution of
the Riccati equation. But additional simulations with various other
choices of functions $\omega(t), m(t)$ suggest that the resulting
momentum distribution defined as above in Eq.~(\ref{momdistr}) has a
step-like form. A formation of an effective momenta scale is
associated with asymptotic emergence of microcanonical-type
distribution.

The Luttinger liquid expression for the $g_{1}$-correlation function
is a low-energy approximation for the true behavior of the
correlation function. However, in the non-equilibrium dynamics we
excite the whole spectrum and therefore the result for our
time-dependent theory based on the exact equilibrium theory may
appear to be different from the one based on the low-energy
approximation. In what follows we demonstrate that the long-time
behavior of the momentum distribution of the time-dependent system
has a bounded support in momentum space. Our arguments can be
applied to any exactly-solvable models.

Suppose the $g_{1}(x,y)$-correlation function is defined as a
ground-state correlator of some field operators
$\Psi(x),\Psi^{\dag}(x)$: $g_{1}(x,y)=\langle
\Psi^{\dag}(x)\Psi(y)\rangle$. We also assume that the matrix
elements of the operator $\Psi(x)$ in the eigenbasis of the
equilibrium problem are known. This implies that the form-factors
$F(\{\lambda\},\{\mu\})=\langle\{\lambda\}|\Psi(0)|\{\mu\}\rangle$
and the norms of the eigenstates $|\lambda\rangle$ and $|\mu\rangle$
are known. Here $\{\mu\},\{\lambda\}$ are the sets of numbers which
characterize the eigenstates of a system of size $2l$. In particular
these numbers can correspond to the solutions of the Bethe ansatz
equations in the exactly-solvable problems. We also assume space-
and time-translation invariance. Therefore the time-dependent
$g_{1}$ function can be expanded as follows
\beq
g_{1}^{(\Phi)}(x,t;0,0)=\sum_{\{\mu\}}\exp[i(E_{\lambda}-E_{\mu})t-i(P_{\lambda}-P_{\mu})x]\frac{|F(\{\mu\},\{\lambda\})|^{2}}{||\lambda||^{2}||\mu||^{2}},
\eeq
where $E_{\lambda}$ and $P_{\lambda}$ are, respectively, energy and
momentum of the state $|\lambda\rangle$. We assume also that the set
$\{\lambda\}$ corresponds to the ground state. Introducing the
coordinates $\xi=x-y$ and $\eta=x+y$, the momentum distribution of
the time-dependent system (we take for simplicity equal-time
correlation function) can be written as
\beq
n(p,t)&=&L(t)\sum_{\{\mu\}}\int_{-l/L(t)}^{l/L(t)}d\xi\int_{-l/L(t)}^{l/L(t)}d\eta\\
&&\quad\times\exp\left[-i\left((P_{\mu}-P_{\lambda})+L(t)p+F(t)L^{2}(t)\eta)\xi\right)\right]
\frac{|F(\{\mu\},\{\lambda\})|^{2}}{||\lambda||^{2}||\mu||^{2}}.\notag
\eeq
The $\xi$-integration can be done easily, while after the
$\eta$-integration we obtain
\beq
n(p,t)=\frac{2l}{F(t)L(t)}\sum_{\{\mu\}}\sum_{\sigma=\pm}\left(\sigma
\mbox{Si}(x_{\sigma})\right)
\frac{|F(\{\mu\},\{\lambda\})|^{2}}{||\lambda||^{2}||\mu||^{2}},
\eeq
where $\mbox{Si}(z)$ is a sine-integral and
$x_{\sigma}=(P_{\mu}-P_{\lambda})l/L(t)+lp+\sigma F(t)l^{2}$. The
integrand is essentially proportional to
$F^{-1}(t)\sin[(P_{\mu}-P_{\lambda}+pL(t))/L(t)]/[(P_{\mu}-P_{\lambda}+pL(t))/L(t)]$
and gives a main contribution to the sum when the momentum transfer
is equal to $pL(t)$.

\subsection{Evaluation of the momentum distribution function in 2D for the uniform system.}

Here we evaluate the momentum distribution function for the 2D Bose
gas below the BKT transition. We consider a system with
time-dependent parameters and assume the validity of the long
wavelength approximation.

\begin{figure}[ht]
\begin{center}
\includegraphics[width=\figwa\textwidth,angle=0]{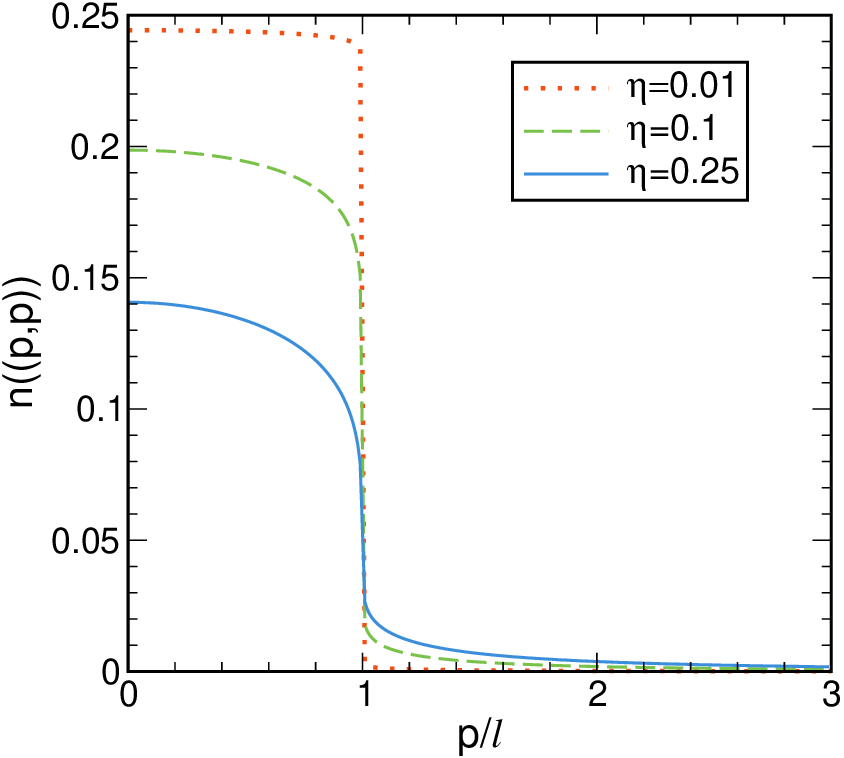}
\caption{
Normalized momentum distributions $n(\bp)$ in the asymptotic-time limit for
temperatures below the BKT-transition \label{fig:np}
}
\end{center}
\end{figure}
According to (\ref{momdistr2}) the momentum distribution in 2D is given
by
\beq
n({\bf p},t)&=&L^{2}(t)\int_{||\br||\leq l/L(t)} \int_{||\br '||\leq
l/L(t)}d\br d\br'g_{1}^{(\Phi)}(\br,\br';0)\\
&&\quad\times\exp[-i
F(t)L^{2}(t)({\bf r}^{2}-\br'^{2})-iL(t){\bf p}\cdot(\br-\br')]\notag.
\eeq
where the integration is restricted to a finite surface of the order
of $(2l/L(t))^{2}$. We  choose the density matrix in the scaling
form corresponding to temperatures below the BKT transition,
\beq
g_{1}^{(\Phi)}(\br,\br';0)
=\rho_{0}\left(\frac{\xi_T}{|\br-\br'|}\right)^{\eta}
\eeq
where $\rho_{0}$ is the density and $\eta
=m_0T/(2\pi\hbar^{2}\rho_{s}(T))$ ($\eta_{BKT}=1/4$). Introducing
the center of mass and relative coordinates
\beq
x=x_{1}-x_{2},\qquad y=y_{1}-y_{2},\qquad
X=\frac{x_{1}+x_{2}}{2},\qquad Y=\frac{y_{1}+y_{2}}{2},
\eeq
and assuming the integration from $-l$ to $l$ we rewrite the
momentum distribution as
\beq
n(\bp,t)\!&=&\!L^{2}(t)\rho_{0}u^{\eta}\!\int_{-l/L(t)}^{l/L(t)}\!\!\!dx\!\int_{-l/L(t)}^{l/L(t)}\!\!\!dy\!\int_{-l/L(t)}^{l/L(t)}
\!\!\!dX\!\int_{-l/L(t)}^{l/L(t)}\!\!\!dY\\&&\quad\times
\frac{\exp[i2F(t)L^{2}(t)(xX+yY)+iL(t)(p_{x}x+p_{y}y)]}{(x^{2}+y^{2})^{\eta/2}},\notag
\eeq
($\bp=(p_x,p_y)$), which after integration over $X$ and $Y$ and changing variables to
\beq
x&\rightarrow&\tilde{x}=2lF(t)L(t)x\equiv x,\notag\qquad
y\rightarrow\tilde{y}=2lF(t)L(t)y\equiv y,\\
A&=&2l^{2}F(t),\qquad p_{x,y}\rightarrow
\tilde{p}_{x,y}=\frac{p_{x,y}}{2lF(t)}
\eeq
takes the following form
\beq\label{np2}
n(\bp,t)&=&\frac{4l^{2}\rho_{0}\xi_T^{\eta}}{(2lF(t)L(t))^{2-\eta}}\int_{-A}^{A}dx\int_{-A}^{A}dy
\frac{\sin(x)}{x}\frac{\sin(y)}{y}\frac{e^{i\tilde{p}_{x}x+i\tilde{p}_{y}y}}{[x^{2}+y^{2}]^{\eta/2}}\notag\\
&=&\sum_{\alpha,\beta=\pm}I_{\alpha\beta},\\
I_{\alpha,\beta}&=&\frac{l^{2}\rho_{0}\xi_T^{\eta}}{(2F(t)L(t))^{2-\eta}}\int_{-A}^{A}dx\int_{-A}^{A}dy
\frac{\sin[x(1+\alpha \tilde{p}_{x})]}{x}\frac{\sin[y(1+\beta\tilde{p}_{y})]}{[x^{2}+y^{2}]^{\eta/2}y}\notag
\eeq
Now, using the integral
$\int_{0}^{\infty}e^{-px^{\mu}}=p^{-1/\mu}\Gamma(1+\frac{1}{\mu})$
we rewrite
\beq
\frac{1}{[x^{2}+y^{2}]^{\eta/2}}=\frac{1}{\Gamma(1+\eta/2)}\int_{0}^{\infty}e^{-(x^{2}+y^{2})t^{2/\eta}}dt
\eeq
and substitute back to Eq.~(\ref{np2}). Then the $x$ and $y$
integrals are separated now and can be performed using
\beq
\int_{-\infty}^{\infty}\frac{e^{-x^{2}t^{2/\eta}}\sin(C x)}{x}dx
=\pi\, \mbox{erf}\left(\frac{|B|}{2t^{1/\eta}}\right)\mbox{sign}(C)
\eeq
where we assume that the integration region can be effectively
extended to infinity. This in particular is justified for large
times when $F(t)$ is a growing function of time or for large $l$ for
arbitrary time. We therefore end up with the following integral
\beq
n(\bp,t)&=&\frac{\pi^{2}l^{2}\rho_{0}\xi_T^{\eta}}{(2lF(t)L(t))^{2-\eta}\Gamma(1+\frac{\eta}{2})}
\sum_{\alpha,\beta=\pm}\notag
\mbox{sign}(1+\alpha\tilde{p}_{x})\mbox{sign}(1+\beta\tilde{p}_{y})\\
&&\quad\times\int_{0}^{\infty}
\mbox{erf}(|1+\alpha\tilde{p}_{x}|/2t^{1/\eta})\mbox{erf}(|1+\beta\tilde{p}_{y}|/2t^{1/\eta})dt
\eeq
which after the change of variables is transformed into the form
\beq
n(\bp,t)=\sum_{\alpha,\beta=\pm}{\cal
N}_{\alpha,\beta}\int_{0}^{\infty}\frac{\mbox{erf}(|a_{\alpha}|u)\mbox{erf}(|b_{\beta}|u)}{u^{\eta+1}}du
\eeq
where
\beq
{\cal
N}_{\alpha,\beta}&=&\frac{\pi^{2}l^{2}\rho_{0}\xi_T^{\eta}(-\eta)\mbox{sign}(a_{\alpha})\mbox{sign}(b_{\beta})}
{(2lF(t)L(t))^{2-\eta}\Gamma(1+\frac{\eta}{2})2^{\eta}}\\
a_{\alpha}&=&1+\alpha \tilde{p}_{x},\qquad b=1+\beta_{\beta}
\tilde{p}_{y}
\eeq
The last integral is equal to
\beq\label{Iab}
\tilde{I}(a_{\alpha},b_{\beta})&=&\frac{i}{2\pi}\left(|a_{\alpha}|^{\eta}B(-\frac{b^{2}_{\beta}}{a^{2}_{\alpha}},\frac{1}{2},\frac{\eta}{2})-
i^{\eta}|b_{\beta}|^{\eta}B(-\frac{b^{2}_{\beta}}{a^{2}_{\alpha}},\frac{1-\eta}{2},\frac{\eta}{2})\right)\notag\\&&+\quad
\frac{|b_{\beta}|^{\eta}\sqrt{\pi}\sec(\frac{\pi\eta}{2})}{\eta\Gamma(\frac{1+\eta}{2})}
\eeq
where $B(.,.)$ is the Euler beta-function.
So, finally we obtain
\beq\label{2d}
n(\bp,t)&=&\frac{\pi^{2}l^{2}\rho_{0}\xi_T^{\eta}(-\eta)}
{(2lF(t)L(t))^{2-\eta}\Gamma(1+\frac{\eta}{2})2^{\eta}}\notag\\&&\quad\times\sum_{\alpha,\beta=\pm}\mbox{sign}(1+\alpha
\tilde{p}_{x})\mbox{sign}(1+\beta
\tilde{p}_{y})\tilde{I}(a_{\alpha},b_{\beta}),
\eeq
where $\tilde{I}(a_\alpha,b_\beta)$ is given in Eq.~(\ref{Iab}),
$a_{\alpha}\equiv 1+\alpha p_{x}/(2lF(t))$ and $b_{\alpha}\equiv
1+\beta p_{y}/(2lF(t))$. We also introduced
$\tilde{p}_{x,y}=p_{x,y}/2lF(t)$.

In fig. \ref{fig:np} we plot the asymptotic behavior of the momentum distributions for various values of $\eta$.
Similarly to the one-dimensional case we find a step-like distribution which is smeared off when the BKT-transition is approached.


\section*{References}
\providecommand{\newblock}{}

\end{document}